\documentclass[apj]{emulateapj-rtx4}

\usepackage{calc}
\shorttitle{Disappearance of the Progenitor of iPTF13bvn}
\shortauthors{Folatelli et al.}

\begin{document}

\title{Disappearance of the Progenitor of Supernova iPTF13bvn}

\author{%
  Gast\'on Folatelli\altaffilmark{1,2},
  Schuyler D.\ Van Dyk\altaffilmark{3},
  Hanindyo Kuncarayakti\altaffilmark{4,5},
  Keiichi Maeda\altaffilmark{6,2},
  Melina C.\ Bersten\altaffilmark{1,2},
  Ken'ichi Nomoto\altaffilmark{2,7},
  Giuliano Pignata\altaffilmark{8,4},
  Mario Hamuy\altaffilmark{5,4},
  Robert M.\ Quimby\altaffilmark{9,2},
  WeiKang Zheng\altaffilmark{10},
  Alexei V.\ Filippenko\altaffilmark{10},
  Kelsey I. Clubb\altaffilmark{10},
  Nathan Smith\altaffilmark{11},
  Nancy Elias-Rosa\altaffilmark{12},
  Ryan J.\ Foley\altaffilmark{13,14},
  and
  Adam A.\ Miller\altaffilmark{15,16}
}

\altaffiltext{1}{Facultad de Ciencias Astron\'omicas y
  Geof\'{\i}sicas, Universidad Nacional de La Plata, Paseo del Bosque
  S/N, B1900FWA La Plata; Instituto de Astrof\'isica de La Plata
  (IALP), CONICET, Argentina}
\altaffiltext{2}{Kavli Institute for the Physics and Mathematics of
  the Universe (WPI), The University of Tokyo, Kashiwa, Chiba
  277-8583, Japan; gaston.folatelli@ipmu.jp}
\altaffiltext{3}{IPAC/Caltech, Mailcode 100-22, Pasadena, CA 91125, USA}
\altaffiltext{4}{Millennium Institute of Astrophysics (MAS), Santiago,
  Chile}   
\altaffiltext{5}{Departamento de Astronom\'ia, Universidad de Chile,
  Casilla 36-D, Santiago, Chile}  
\altaffiltext{6}{Department of Astronomy, Kyoto University,
  Kitashirakawa-Oiwake-cho, Sakyo-ku, Kyoto 606-8502, Japan} 
\altaffiltext{7}{Hamamatsu Professor}
\altaffiltext{8}{Departamento de Ciencias Fisicas, Universidad Andres
Bello, Avda. Republica 252, Santiago, Chile} 
\altaffiltext{9}{Department of Astronomy, San Diego State University,
  5500 Campanile Drive, San Diego, CA 92182-1221, USA}  
\altaffiltext{10}{Department of Astronomy, University of California,
Berkeley, CA 94720-3411, USA} 
\altaffiltext{11}{Steward Observatory, University of Arizona, 933
  N. Cherry Ave., Tucson, AZ 85721, USA} 
\altaffiltext{12}{INAF-Osservatorio Astronomico di Padova, Vicolo
  dell'Osservatorio 5, I-35122 Padova, Italy} 
\altaffiltext{13}{Astronomy Department, University of Illinois at
  Urbana-Champaign, 1002 W.\ Green Street, Urbana, IL 61801, USA}
\altaffiltext{14}{Department of Physics, University of Illinois at
  Urbana-Champaign, 1110 W.\ Green Street, Urbana, IL 61801, USA}
\altaffiltext{15}{Jet Propulsion Laboratory, 4800 Oak Grove Drive, MS
  169-506, Pasadena, CA 91109, USA}
\altaffiltext{16}{Hubble Fellow}

\begin{abstract}
\noindent Supernova (SN) iPTF13bvn in NGC~5806 was the first Type~Ib SN
to have been tentatively associated with a progenitor in 
pre-explosion images. 
We performed deep ultraviolet (UV) and optical {\em Hubble Space
  Telescope} observations of the SN site
$\sim740$ days after explosion. We detect an object in the optical
bands that is fainter than the pre-explosion object. This dimming is
likely not produced by dust absorption in the ejecta;
thus, our finding confirms the
connection of the progenitor candidate with the SN. The object in our
data is likely dominated by the fading SN, implying that
the pre-SN flux is mostly due to the progenitor. We compare our
revised pre-SN photometry with previously proposed
models. Although binary progenitors are favored, models need to be
refined. In particular, to comply with our deep UV detection
limit, any companion star must be less luminous than
a late-O star or substantially obscured by newly formed dust. A
definitive progenitor characterization will require further
observations to disentangle the contribution of a much fainter SN and
its environment.  
\end{abstract}


\section{INTRODUCTION}
\label{sec:intro}

\noindent The stellar origin of hydrogen-free core-collapse
supernovae (SNe) remains unknown chiefly owing to the lack of detections
of progenitor stars in pre-explosion images \citep{Eldridge13}. The only firm
progenitor candidate found thus far is that of iPTF13bvn, a Type Ib
supernova (SN~Ib) in the galaxy NGC~5806 \citep{Cao13}. If confirmed,
this case can provide important clues about the mechanisms of envelope
removal among massive stars. One proposed mechanism for very massive
stars ($M_{\mathrm{ZAMS}}>25$\,M$_\odot$) is strong stellar winds
leading to Wolf-Rayet (WR) progenitors \citep{Heger03}, but such massive
progenitors are difficult to reconcile with the large fraction of 
stripped-envelope explosions \citep{Smith11} and their low ejecta masses
\citep{Drout11,Dessart11,Hachinger12}. The more common alternative is
mass transfer in close binary systems
\citep{Shigeyama90,Podsiadlowski92}, allowing less-massive stars to
lose their envelopes \citep[e.g.,][]{Benvenuto13}. The 
relative incidence of each type of progenitor is still unknown. 

\citet{Cao13} used pre-explosion {\em Hubble Space Telescope (HST)}
imaging, together with early-time SN observations, to suggest
that the progenitor of iPTF13bvn was a compact WR star. \citet{Groh13b}
found that the pre-explosion photometry could be fit with models of
single WR stars in the initial mass range of 31--35\,M$_\odot$. 
However, from hydrodynamical light-curve modeling,
\citet{Bersten14} inferred a low pre-SN mass ($\sim3.5$\,M$_\odot$) 
and disfavored a massive WR star; instead, they presented
close binary system models that could explain the pre-explosion
photometry and the progenitor mass and radius
derived from the SN observations. 
By analyzing the light curves, \citet{Fremling14}
and \citet{Srivastav14} also disfavored a massive WR
progenitor. \citet{Kuncarayakti15} argued for a 
low-mass progenitor based on the strength of oxygen and calcium lines
in the late-time spectrum. Subsequently, \citet{Eldridge15} (E15,
hereafter) revised the 
pre-explosion photometry and found the progenitor to be 
brighter than measured by \citet{Cao13}; see also
our measurements in Section~\ref{sec:obs}. With the new magnitudes
and using binary evolution calculations, E15 argued in
favor of the binary scenario. The same conclusion was found by
\citet{Kim15}. Assuming the binary configurations discussed by
  \citet{Bersten14}, \citet{Hirai15} simulated the effect of the SN shock
  on the companion star. They found that the companion could bloat and
  thus evolve to a red-supergiant structure on a timescale
of a few years after explosion. We note, however, that similar shock
simulations performed by \citet{Liu15}, although for different
companion masses, do not predict such a post-explosion evolution.

To better determine the progenitor's nature, 
the SN site had to be reobserved after the ejecta faded
enough to probe the disappearance of the pre-SN object. In this work
we present new {\em HST} observations that reveal a decrease in flux
relative to the pre-SN observations, confirming that this is the first
SN~Ib with a progenitor detection\footnote{In the final stages of preparing 
this manuscript, \citet{Eldridge16} independently reported the disappearance
of the progenitor candidate using some of the data presented herein.}.
In Section~\ref{sec:obs} we
describe the observations and photometry methods.
The new data are analyzed in Section~\ref{sec:prog} 
to constrain the progenitor nature. We present our conclusions
in Section~\ref{sec:concl}.

\section{OBSERVATIONS AND PHOTOMETRY}
\label{sec:obs}                 %

\begin{figure*}[htpb] 
\begin{center}
\includegraphics[width=0.32\textwidth]{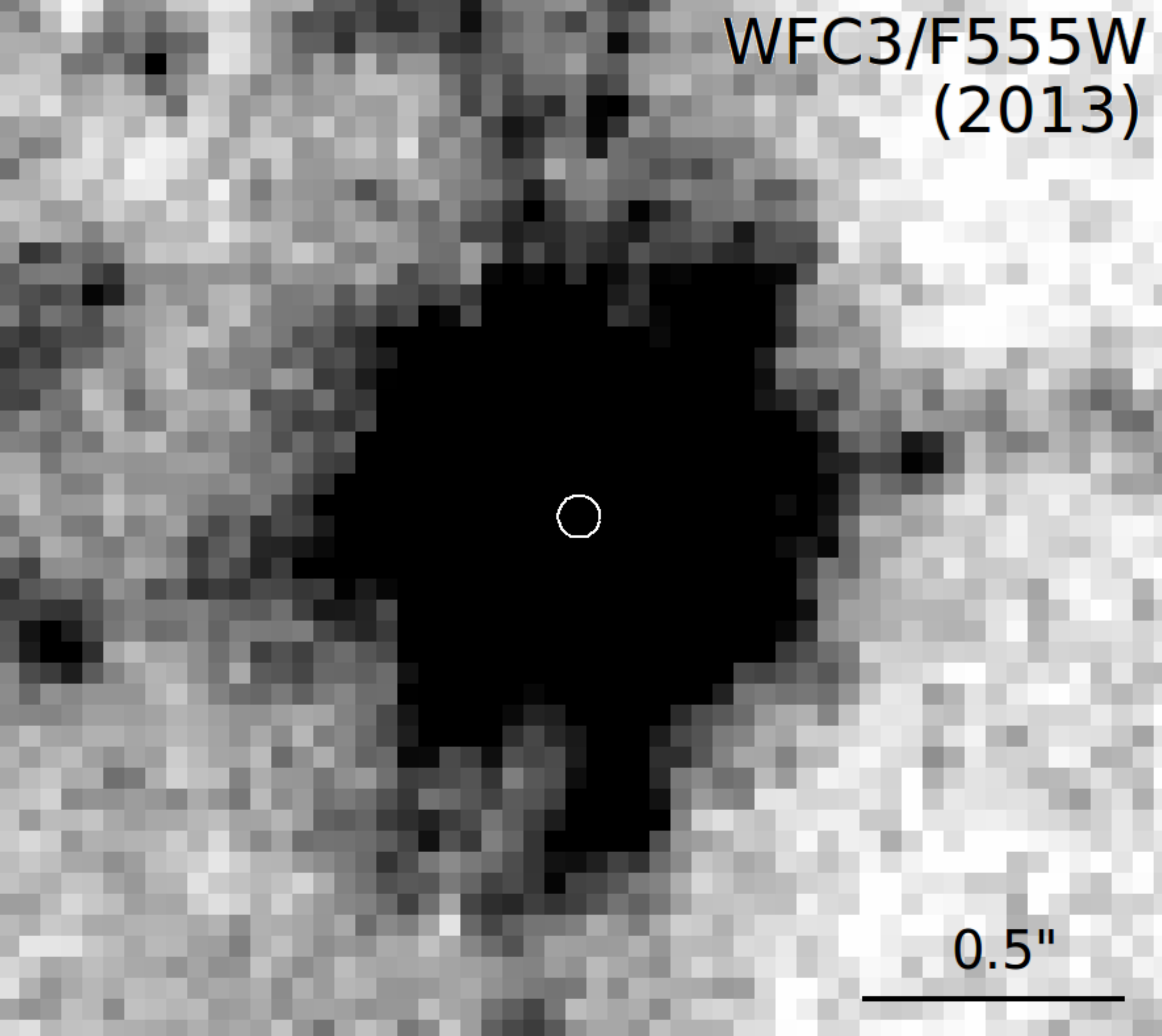}\hspace{0.01\textwidth}\includegraphics[width=0.32\textwidth]{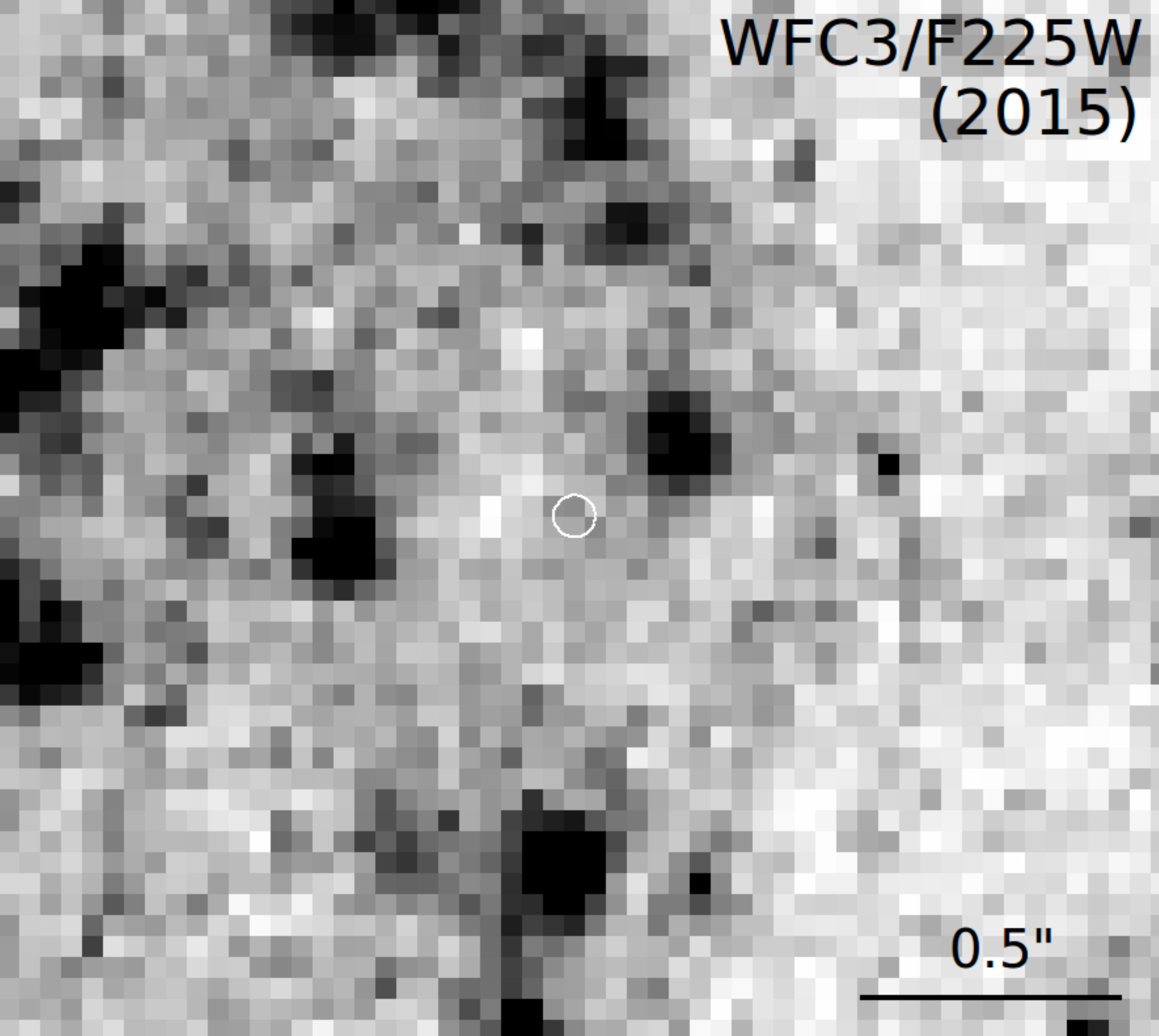}\hspace{0.33\textwidth}

\vspace{0.01\textheight}

\includegraphics[width=0.32\textwidth]{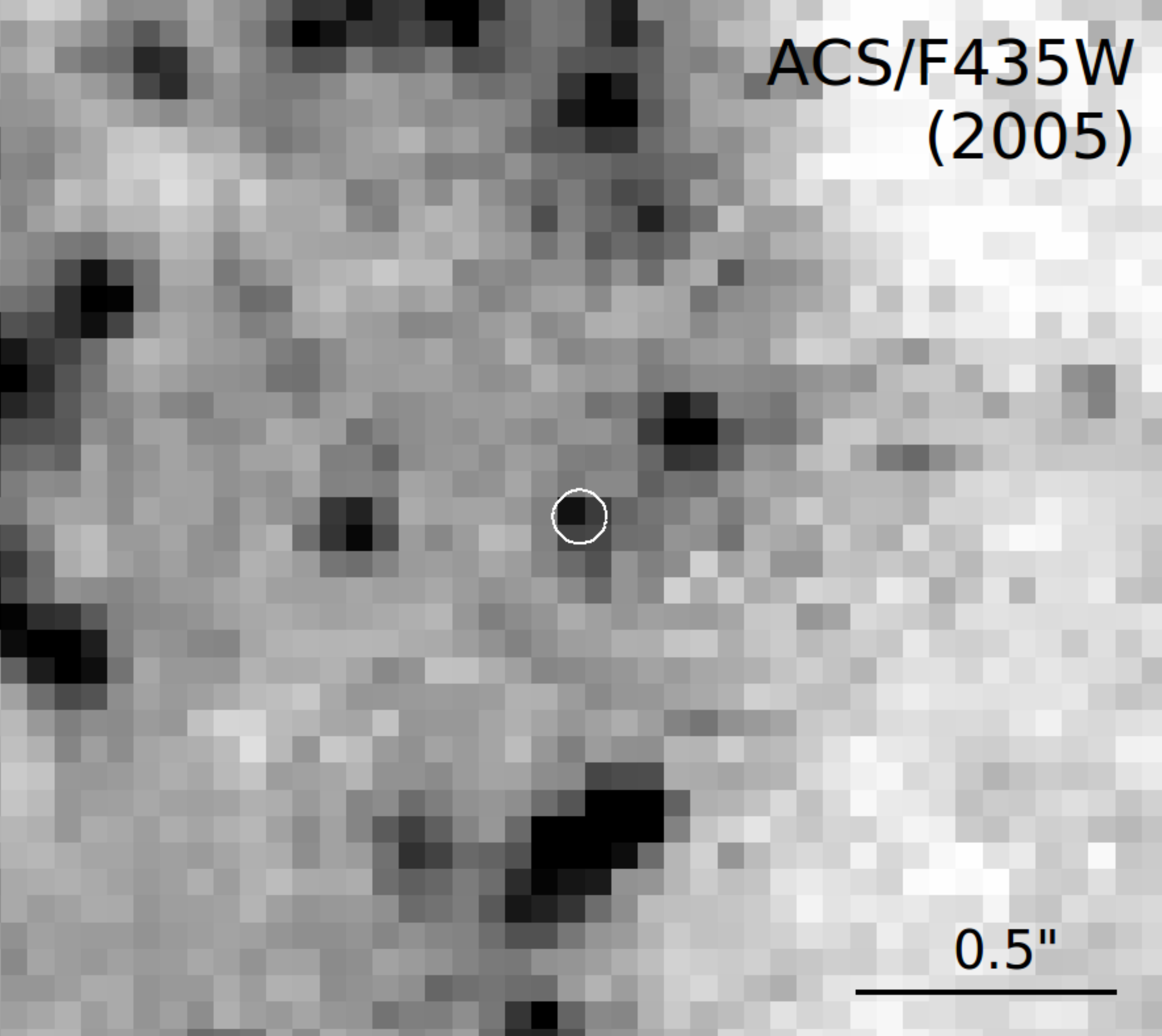}\hspace{0.01\textwidth}\includegraphics[width=0.32\textwidth]{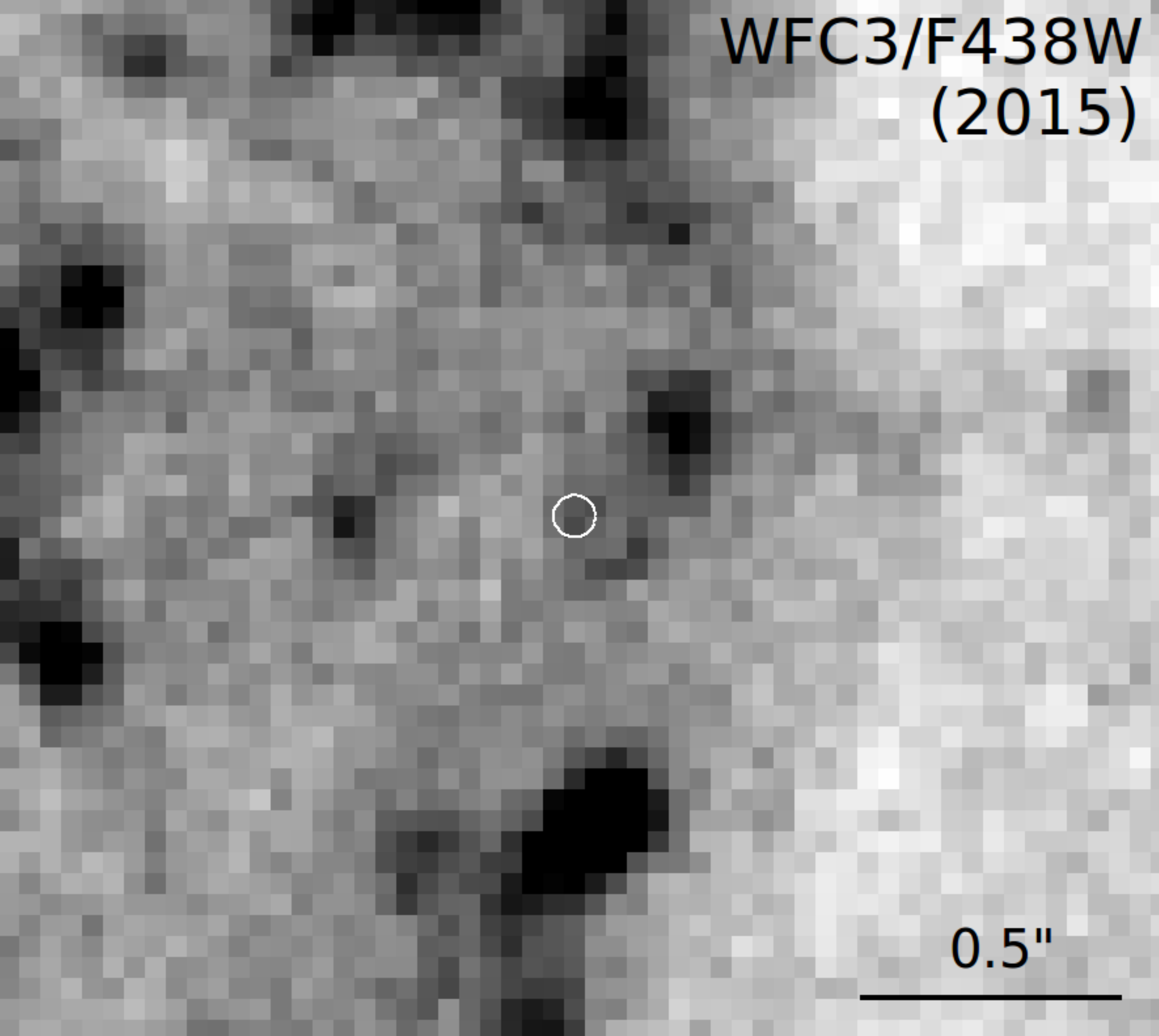}\hspace{0.01\textwidth}\includegraphics[width=0.32\textwidth]{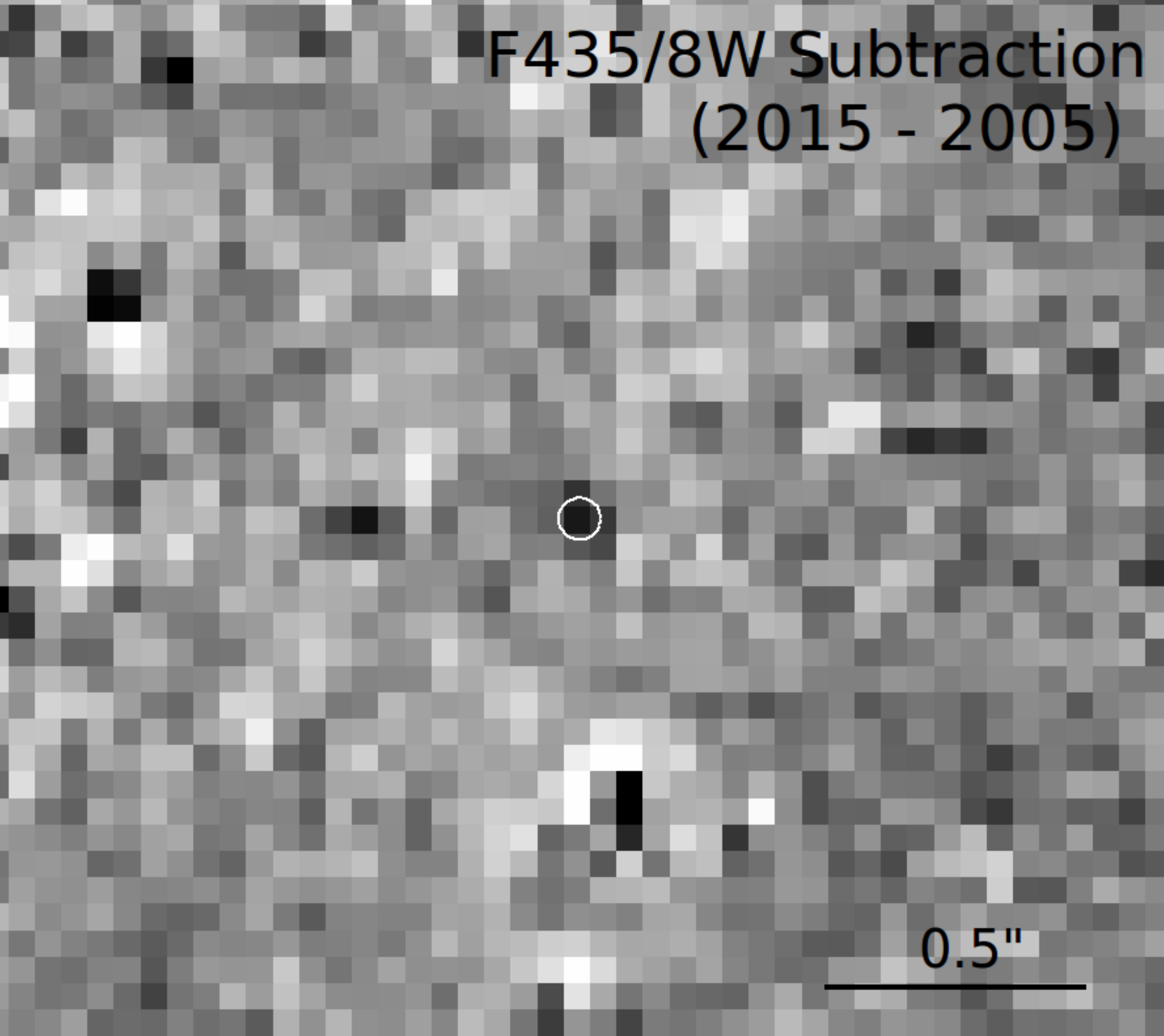}

\vspace{0.01\textheight}

\includegraphics[width=0.32\textwidth]{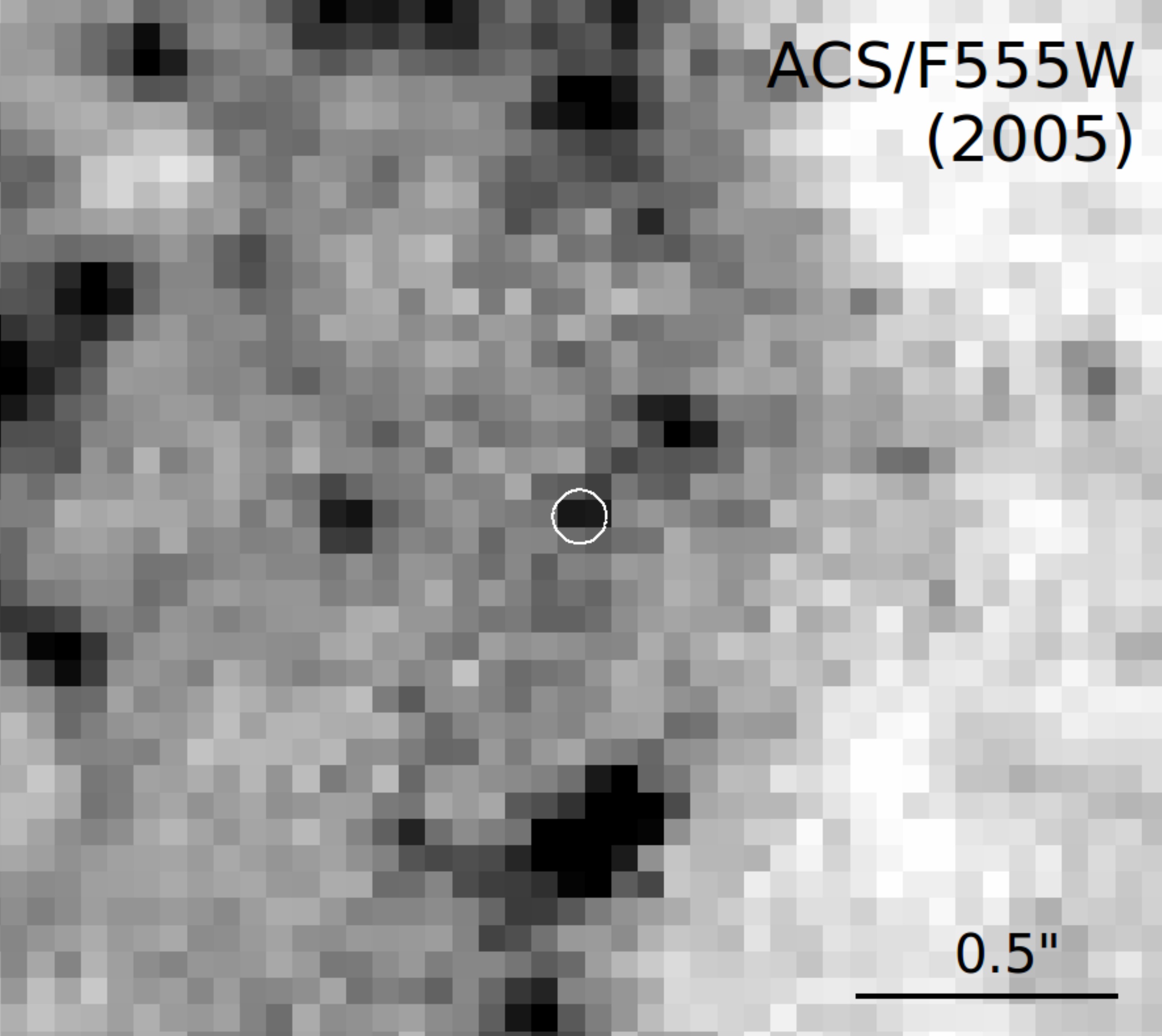}\hspace{0.01\textwidth}\includegraphics[width=0.32\textwidth]{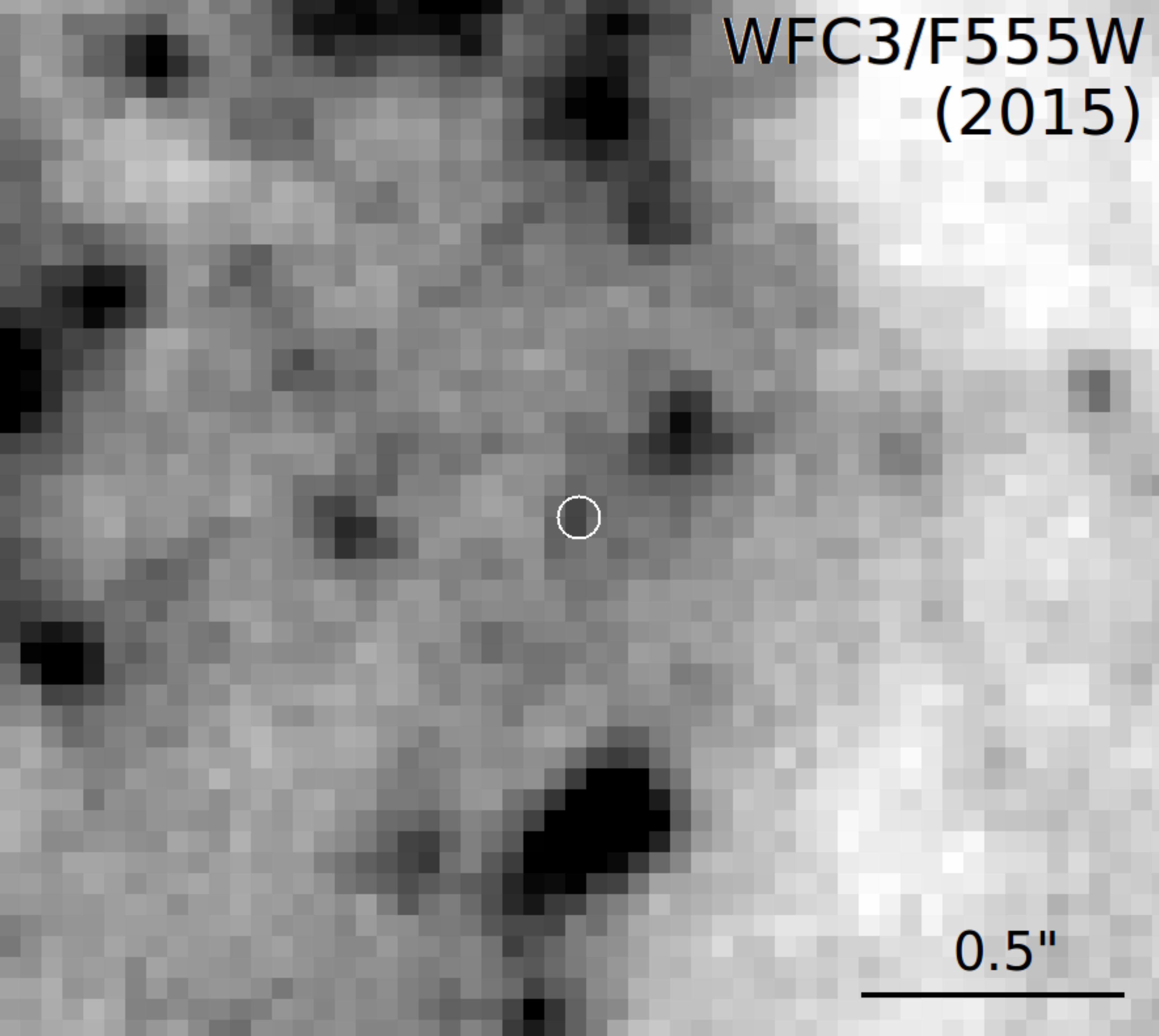}\hspace{0.01\textwidth}\includegraphics[width=0.32\textwidth]{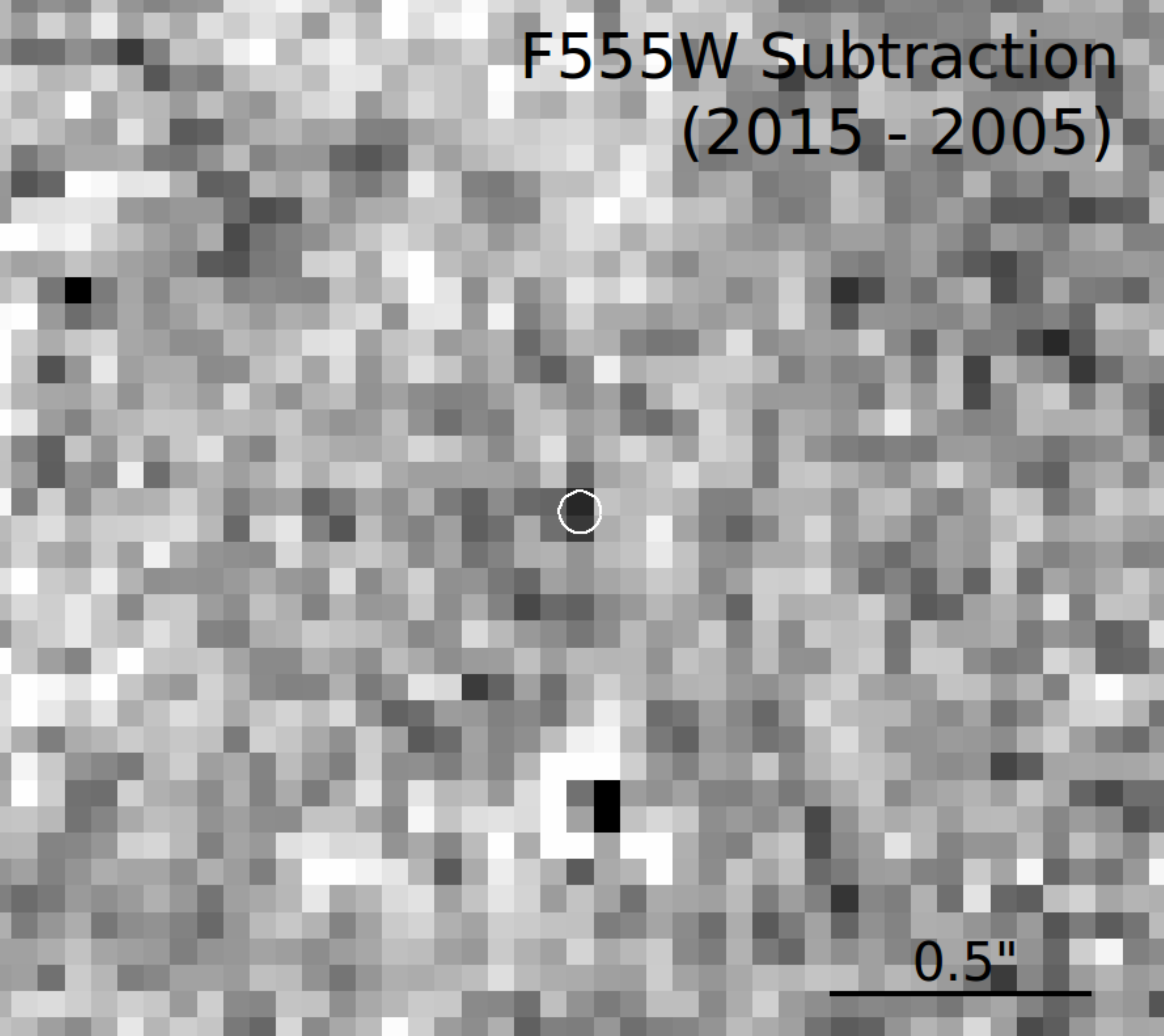}

\vspace{0.01\textheight}

\includegraphics[width=0.32\textwidth]{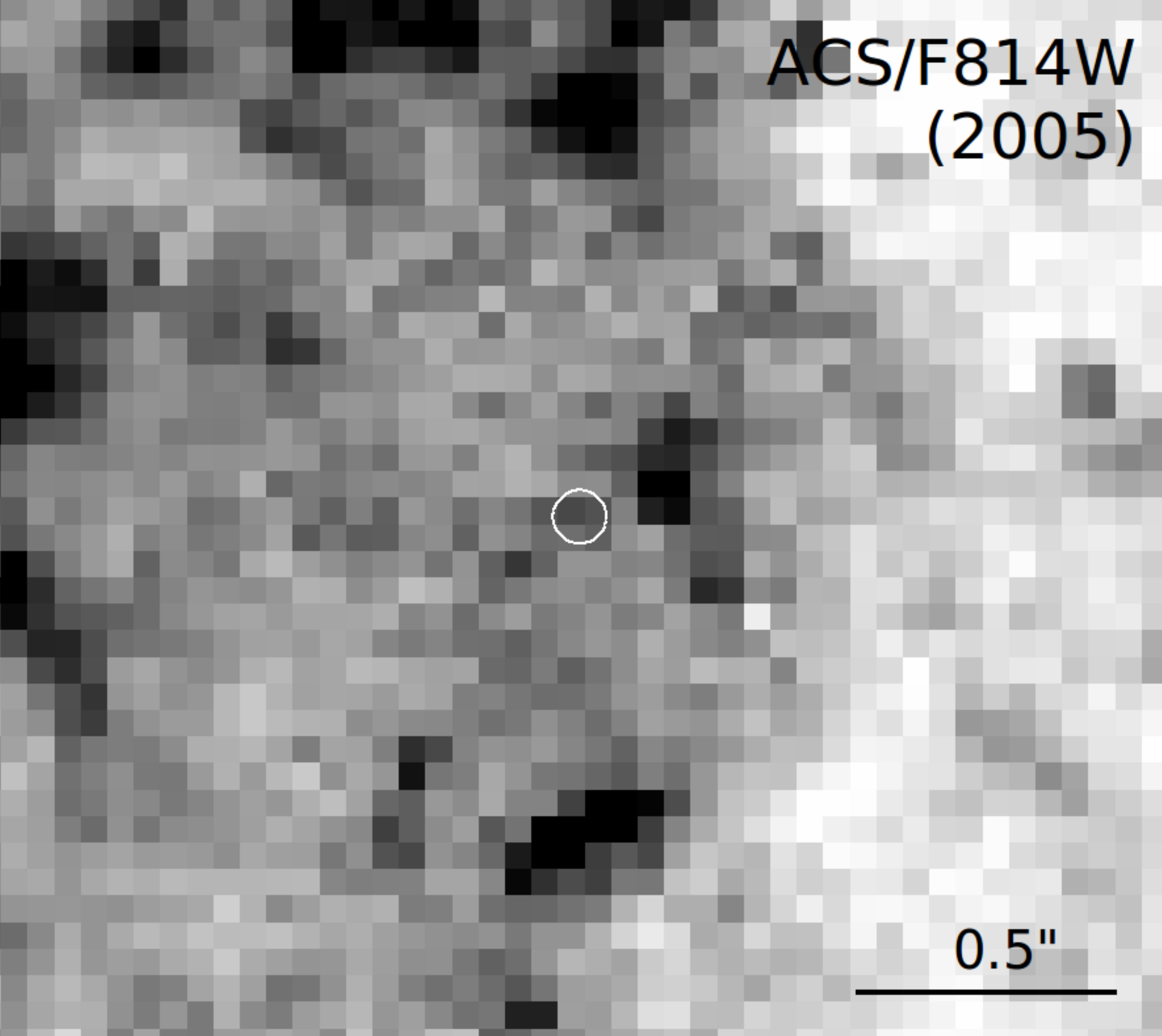}\hspace{0.01\textwidth}\includegraphics[width=0.32\textwidth]{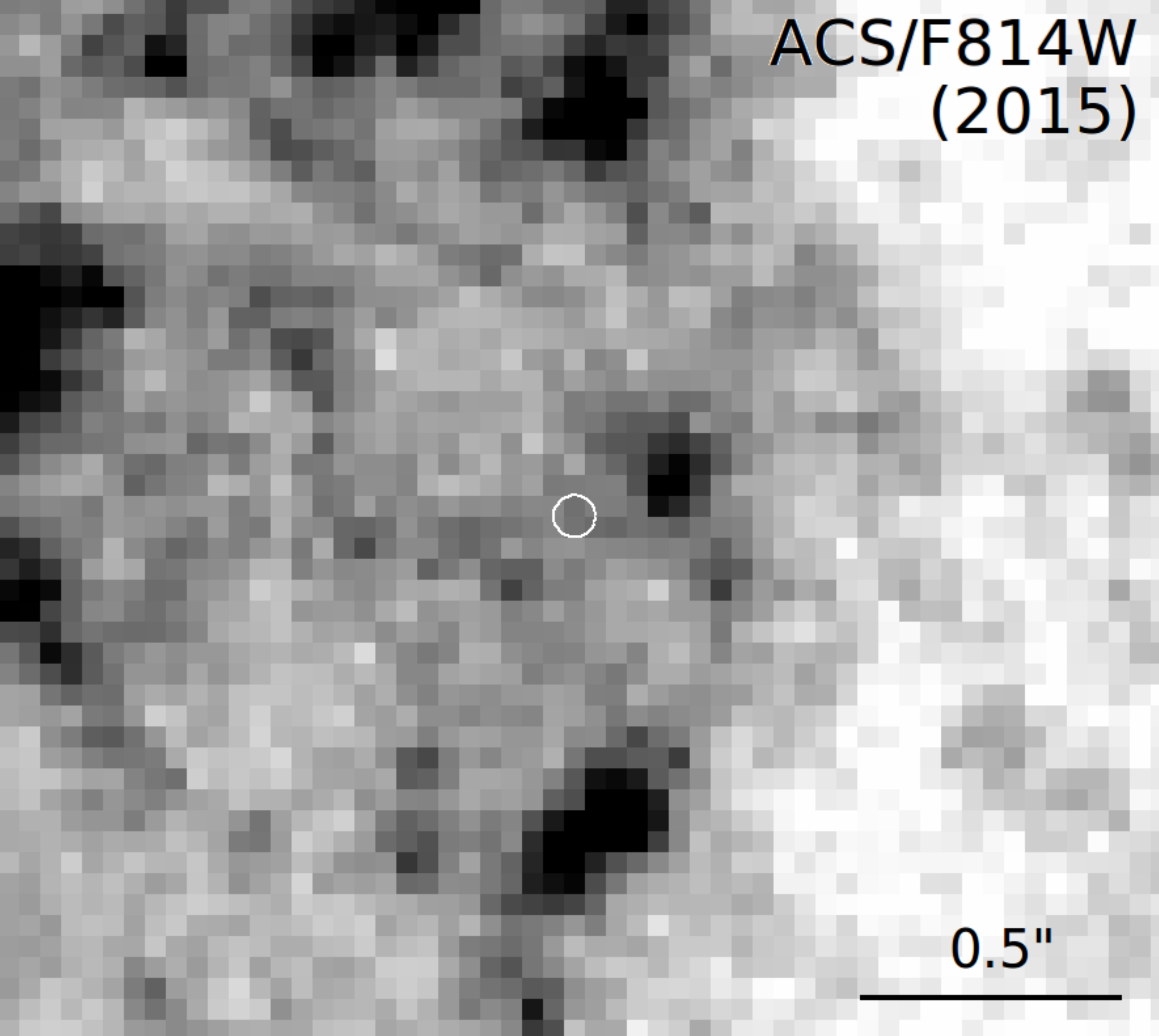}\hspace{0.01\textwidth}\includegraphics[width=0.32\textwidth]{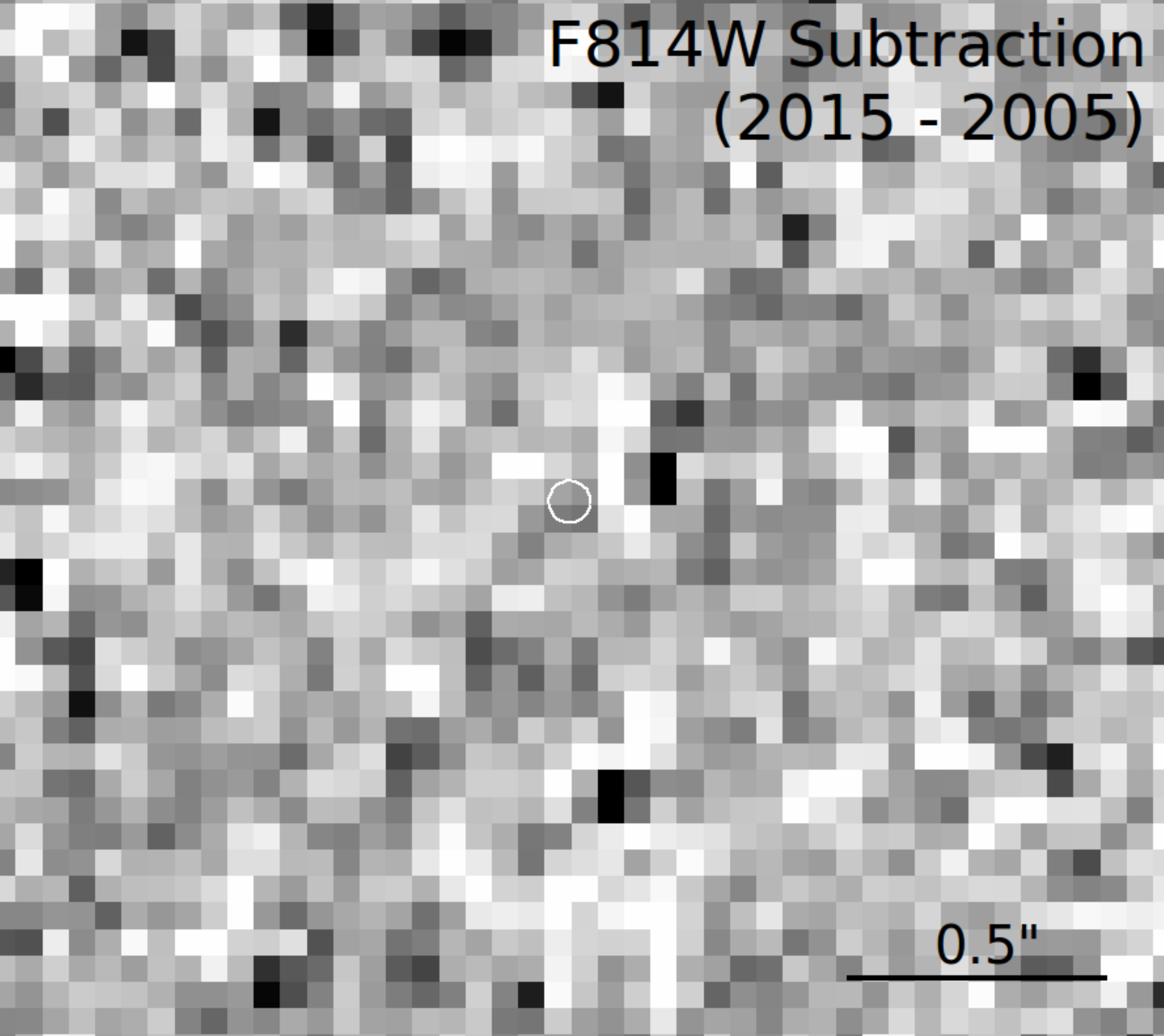}
\caption{{\em HST} images of the site of iPTF13bvn at
  different epochs. {\em Top left:} An image near maximum
  light to locate the SN. {\em Rest of left column:}
  Pre-SN images obtained in 2005. {\em Middle column:} New images
  obtained in 2015. {\em Right column:} Pre-SN minus
  post-SN images. The SN location is shown with a white circle
  of 3$\sigma$ radius. The image scale is indicated. North is up and
  east to the left.
  \label{fig:img}}
\end{center}
\end{figure*}

\noindent We obtained deep imaging of the field of iPTF13bvn 
$\sim740$\,days after explosion using {\em HST} through Cycle 22 programs
GO-13684 
and GO-13822. 
Program GO-13684 was executed between 2015 June 26.37 and 26.60 (UT dates
are used herein) with
the Wide Field Camera 3 (WFC3) UVIS channel; see Table~\ref{tab:img}.
Program GO-13822 comprised observations obtained on 2015 
June 30.63 with WFC3/UVIS (F225W filter)
and on June 30.90 UT with the Advanced Camera for Surveys
(ACS; F814W filter). 
The 2015 images are shown in Figure~\ref{fig:img},
along with the pre-explosion images obtained in 2005 through program
GO-10187 
with ACS. 

The SN location in the pre- and post-explosion images was found
by aligning them relative to a F555W 
image obtained through program GO-12888 
with WFC3/UVIS on 2013 September 
2.37 when the SN was still very bright. 
The registration was done with the AstroDrizzle package employing 20
point sources in common. The resulting precision (root-mean square) 
was 0.081 pixel (32 mas) and 0.090 pixel (36 mas) in the
$X$ and $Y$ axes, respectively. The F435W, F555W, and F814W images from
2015 show an object at the SN location, whereas the F225W image exhibits
no identifiable source at the same place. 

After correcting for charge-transfer-efficiency losses and
  masking of cosmic-ray hits, we performed
photometry on the pre- and post-explosion images using the DOLPHOT
v2.0 package \citep{Dolphin00}. Our measurements are listed in
  Table~\ref{tab:mag} along with others from the literature. The
  pre-explosion photometry given here is roughly consistent with that 
presented by E15 (both their values from DOLPHOT, and the average with
DAOPHOT measurements), and it is 
brighter by 0.3--0.5 mag than what \citet{Cao13} measured. In a recent
work, \citet{Eldridge16} published photometry of the 2015 F438W and
F555W images. Their values are brighter than ours, more significantly
so (by $\sim5\sigma$) in the F555W band. We have 
tried to reproduce their results by changing the parameters in
DOLPHOT, but could not obtain exactly the same values. This may in
part be due to the fact that they apply DOLPHOT on the \_crj files
while we use the \_flc files. Additionally, E15 mention
that they adopt the recommended DOLPHOT parameters. Among these,
the sky-fitting algorithm parameter, FitSky, is suggested to be set as
1 for general purposes in the DOLPHOT v2.0 ACS 
module manual. However, for a crowded field such as that of iPTF13bvn,
the recommendation is to perform the sky fit inside a relatively large
photometry aperture by adopting FitSky=3 \citep{Dalcanton09}, which
was our choice. If we instead use FitSky=1, we  
obtain F555W\,$=26.37 \pm 0.05$ mag, in close agreement with
\citet{Eldridge16}. We also obtain large negative sharpness values, as
mentioned by those authors, which is not seen with FitSky=3. We
conclude that the main source of discrepancy is likely 
the choice of the sky-fitting algorithm, and that for the field of
iPTF13bvn, our procedure is more accurate. Nevertheless, we note that
the main conclusions in the present work would not change if we adopted
the photometry in E15 and \citet{Eldridge16}.

\begin{figure*}[htpb] 
\plottwo{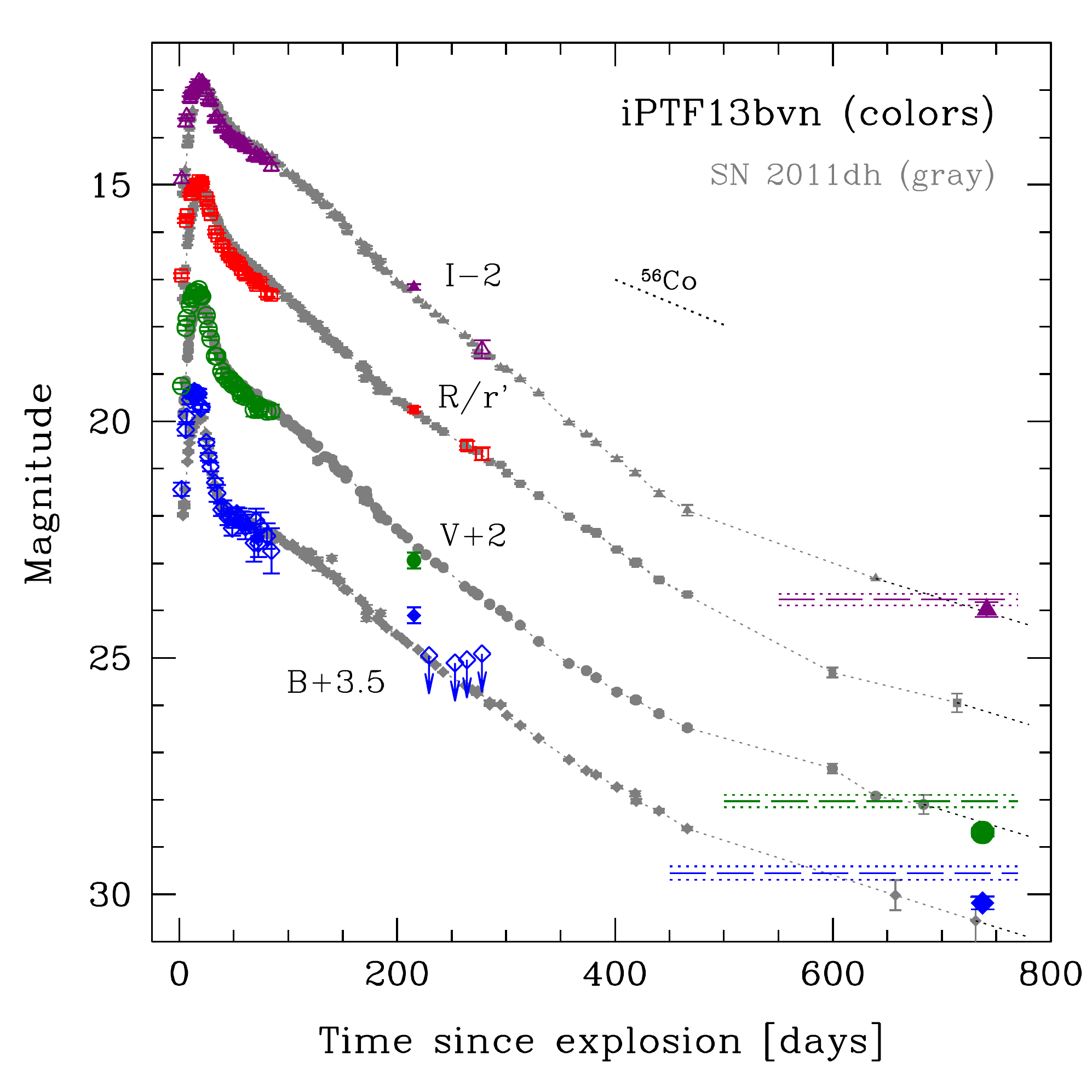}{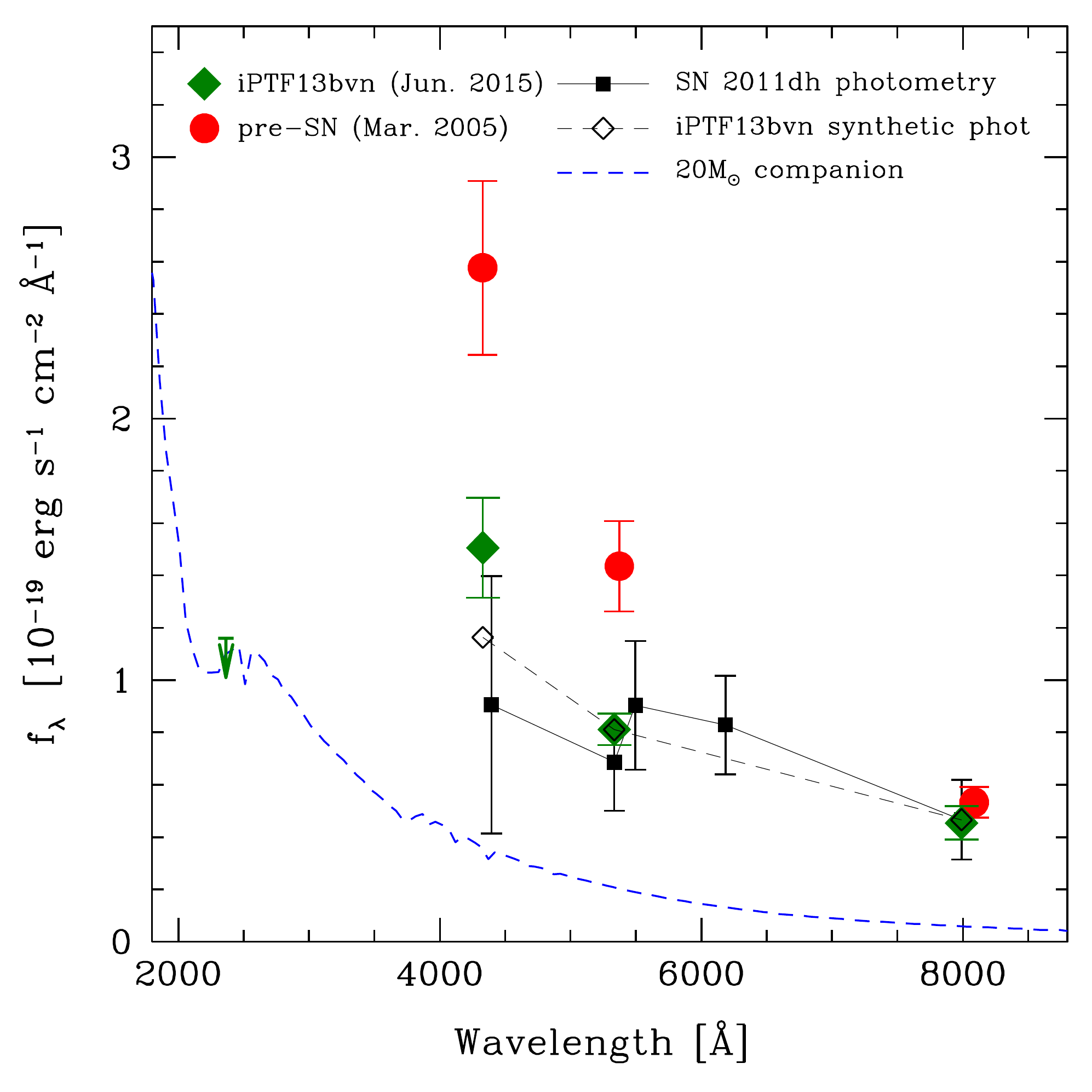}
\caption{{\em Left panel:} $BVRI$ light curves of iPTF13bvn (colored
  symbols) obtained with KAIT 
  (open), Nickel (small filled),
  and {\em HST} (large filled; converted to $BVI$). Detection limits in
  $B$ are indicated with 
  arrows. Gray points connected with dotted lines show the $BVRr'I$
  photometry of SN~2011dh from \citet{Ergon14} and 
  \citet{Ergon15}, scaled to match the distance and extinction of
  iPTF13bvn (see text). Long-dashed horizontal lines indicate the
  magnitude of the pre-explosion object in $BVI$ (1$\sigma$
  uncertainties are indicated with dotted lines). The decline rate
  from $^{56}$Co decay is indicated with a short dotted line. 
  {\em Right panel:} SED of the source
  detected in pre-SN images (red 
  circles) and at $\sim740$ days (green diamonds and green arrow for
  the detection limit in the F225W band). Photometry of SN~2011dh at a
  similar age scaled to the distance and extinction of iPTF13bvn is
  shown with black squares. Synthetic photometry from a nebular spectrum of
  iPTF13bvn scaled to the observed F555W flux is shown with open
  diamonds. The dashed line shows the 20\,M$_\odot$ O-type star
    (``Ostar1'') of \citet{Kim15}, which represents an accreting
    companion star at the moment of the SN explosion. 
  \label{fig:lcssed}}
\end{figure*}

Figure~\ref{fig:lcssed} (left panel) shows the {\em HST} magnitudes
converted to $BVI$. The conversion was calculated using synthetic
photometry from the 306\,day spectrum of \citet{Kuncarayakti15}, which
gave $B-\mathrm{F438W}=0.06$ mag, $V-\mathrm{F555W}=-0.03$ mag, and 
$I-\mathrm{F814W}=-0.05$ mag.

We used the F225W image to compute a detection limit at the SN
location, employing DOLPHOT to add $\sim50,000$ 
artificial stars near the SN site with a uniform distribution of
brightness over 25.4--27.4\,mag. DOLPHOT was run on the
resulting images to recover the artificial sources. A detection limit
of 26.4\,mag was found where the recovery fraction dropped to
50\%, which is roughly equivalent to a 5$\sigma$ detection limit
\citep{Harris90}. 

\begin{deluxetable*}{lccccl}  
\tabletypesize{\small} 
\tablecolumns{6} 
\tablewidth{0pt} 
\tablecaption{Summary of {\em HST} Observations.\label{tab:img}} 
\tablehead{ 
\colhead{UT~Date} & \colhead{Instrument/} & \colhead{Filter}  & \colhead {Exposure} & \colhead{Program} & \colhead{Program}\\
\colhead{} & \colhead{Detector} & \colhead{} & \colhead{(s)} & \colhead{ID} & \colhead{PI} 
}
\startdata 
\multicolumn{6}{c}{Pre-explosion} \\
2005 Mar 10 & ACS/WFC & F435W & 1600 & GO-10187 & Smartt\\
2005 Mar 10 & ACS/WFC & F555W & 1400 & GO-10187 & Smartt\\
2005 Mar 10 & ACS/WFC & F814W & 1700 & GO-10187 & Smartt\\
\multicolumn{6}{c}{Post-explosion} \\
2013 Sep 03 & WFC3/UVIS & F555W & 1200 & GO-12888 & Van Dyk\\
2015 Jun 26 & WFC3/UVIS & F438W & 5720 & GO-13684 & Van Dyk\\
2015 Jun 26 & WFC3/UVIS & F555W & 5610 & GO-13684 & Van Dyk\\
2015 Jun 30 & WFC3/UVIS & F225W & 8865 & GO-13822 & Folatelli\\
2015 Jun 30 & ACS/WFC   & F814W & 2345 & GO-13822 & Folatelli
\enddata 
\end{deluxetable*} 

\begin{deluxetable*}{lccccl}  
\tabletypesize{\small} 
\tablecolumns{6} 
\tablewidth{0pt} 
\tablecaption{Photometry from pre- and post-explosion {\em HST} images.\label{tab:mag}} 
\tablehead{ 
\colhead{UT~Date} & \colhead{F225W} & \colhead{F435W/F438W} & \colhead{F555W} & \colhead{F814W} & \colhead{Source}
}
\startdata 
2005 Mar 10 & $\cdots$ & $26.50$(15) & $26.40$(15) & $26.10$(20) & \citet{Cao13} \\
2005 Mar 10 & $\cdots$ & $25.81$(06) & $25.86$(08) & $25.77$(10) & E15 (DOLPHOT)\\
2005 Mar 10 & $\cdots$ & $25.80$(12) & $25.80$(11) & $25.88$(24) & E15 (Average)\\
2005 Mar 10 & $\cdots$ & $25.99$(14) & $26.06$(13) & $25.82$(12) & This work \\
2015 Jun 26 & $\cdots$ & $26.48$(08) & $26.33$(05) & $\cdots$    & \citet{Eldridge16} \\
2015 Jun 26/30 & $>26.4$\tablenotemark{a} & $26.62$(14) & $26.72$(08) & $26.03$(15) & This work \\
Subtraction (2005--2015) & $\cdots$ & $27.13$(12) & $27.12$(21) & $>27.5$\tablenotemark{a} & This work
\enddata 
\tablecomments{Uncertainties in parentheses in units of
  0.01\,mag.}
\tablenotetext{a}{Limiting magnitude at 50\% detection probability.}
\end{deluxetable*} 

We also obtained $BVRI$ imaging of iPTF13bvn until
$\sim280$ days with the Katzman
Automatic Imaging Telescope \citep[KAIT;][]{Filippenko01} and the 1~m
Nickel telescope at Lick Observatory. Template subtraction was
performed using additional images obtained after the SN faded below
detection. All images were reduced using a custom pipeline
\citep{Ganeshalingam10}. Point-spread-function (PSF) photometry was measured
with the DAOPHOT package \citep{Stetson87}. Apparent magnitudes were
first measured in the KAIT4 natural system and then transformed to the
standard system using local calibrators and color terms as given in
Table~4 of \citet{Ganeshalingam10}. We present the resulting light
curves in Table~\ref{tab:lcs} and 
the left panel of Figure~\ref{fig:lcssed}.

\begin{deluxetable*}{lcccccl}  
\tabletypesize{\small} 
\tablecolumns{7} 
\tablewidth{0pt} 
\tablecaption{KAIT and Nickel photometry of iPTF13bvn.\label{tab:lcs}} 
\tablehead{ 
\colhead{MJD} & \colhead{Phase\tablenotemark{a}} & \colhead{$B$} & \colhead{$V$}  & \colhead{$R$} & \colhead{$I$} & \colhead{Telescope} \\
\colhead{} & \colhead{(days)} & \colhead{(mag)} & \colhead{(mag)} & \colhead{(mag)} & \colhead{(mag)} & \colhead{}
}
\startdata 
$56461.28$ & $2.03$   & $17.94(14)$ & $17.25(07)$ & $16.92(05)$ & $16.88(09)$ & KAIT \\
$56465.22$ & $5.95$   & $16.68(12)$ & $16.02(06)$ & $15.76(05)$ & $15.68(08)$ & KAIT \\
$56466.21$ & $6.94$   & $16.40(11)$ & $15.82(05)$ & $15.64(03)$ & $15.55(04)$ & KAIT \\
$56469.30$ & $10.02$  & $16.04(10)$ & $15.54(04)$ & $15.20(04)$ & $15.18(06)$ & KAIT \\
$56470.27$ & $10.98$  & $15.90(06)$ & $15.39(04)$ & $15.20(04)$ & $15.10(04)$ & KAIT \\
$56471.25$ & $11.95$  & $15.93(06)$ & $15.38(03)$ & $15.18(02)$ & $15.06(03)$ & KAIT \\
$56472.25$ & $12.95$  & $15.95(07)$ & $15.31(03)$ & $15.08(02)$ & $14.97(03)$ & KAIT \\
$56473.21$ & $13.90$  & $15.86(06)$ & $15.26(04)$ & $15.04(03)$ & $14.98(04)$ & KAIT \\
$56474.20$ & $14.90$  & $15.89(09)$ & $15.27(05)$ & $15.01(04)$ & $14.97(04)$ & KAIT \\
$56476.20$ & $16.89$  & $15.90(09)$ & $15.28(04)$ & $14.99(04)$ & $14.86(04)$ & KAIT \\
$56477.20$ & $17.88$  & $15.91(09)$ & $15.21(05)$ & $14.92(03)$ & $14.81(04)$ & KAIT \\
$56479.20$ & $19.87$  & $16.23(07)$ & $15.33(04)$ & $15.00(04)$ & $14.87(05)$ & KAIT \\
$56480.20$ & $20.86$  & $16.19(06)$ & $15.36(03)$ & $14.96(03)$ & $14.83(03)$ & KAIT \\
$56484.25$ & $24.90$  & $16.94(07)$ & $15.76(04)$ & $15.28(03)$ & $15.03(04)$ & KAIT \\
$56486.20$ & $26.84$  & $17.25(08)$ & $16.05(04)$ & $15.48(03)$ & $15.23(04)$ & KAIT \\
$56488.23$ & $28.86$  & $17.46(09)$ & $16.25(04)$ & $15.62(03)$ & $15.25(05)$ & KAIT \\
$56492.22$ & $32.83$  & $17.80(10)$ & $16.62(06)$ & $16.00(05)$ & $15.60(05)$ & KAIT \\
$56494.20$ & $34.80$  & $18.02(18)$ & $16.63(07)$ & $16.07(05)$ & $15.57(05)$ & KAIT \\
$56498.20$ & $38.78$  & $18.37(13)$ & $16.94(05)$ & $16.29(04)$ & $15.81(04)$ & KAIT \\
$56500.23$ & $40.81$  & $18.31(14)$ & $17.03(05)$ & $16.30(04)$ & $15.80(04)$ & KAIT \\
$56504.20$ & $44.75$  & $18.54(15)$ & $17.13(05)$ & $16.46(04)$ & $15.92(04)$ & KAIT \\
$56506.19$ & $46.74$  & $18.52(17)$ & $17.11(05)$ & $16.48(03)$ & $15.95(03)$ & KAIT \\
$56508.19$ & $48.73$  & $18.76(15)$ & $17.22(07)$ & $16.59(05)$ & $16.01(05)$ & KAIT \\
$56510.18$ & $50.71$  & $18.57(18)$ & $17.20(05)$ & $16.61(03)$ & $16.04(04)$ & KAIT \\
$56512.18$ & $52.70$  & $18.45(13)$ & $17.27(06)$ & $16.65(04)$ & $16.11(05)$ & KAIT \\
$56514.18$ & $54.70$  & $18.50(16)$ & $17.32(06)$ & $16.67(04)$ & $16.07(04)$ & KAIT \\
$56516.18$ & $56.68$  & $18.56(16)$ & $17.45(08)$ & $16.77(05)$ & $16.12(05)$ & KAIT \\
$56518.18$ & $58.67$  & $18.64(23)$ & $17.39(09)$ & $16.85(06)$ & $16.13(06)$ & KAIT \\
$56520.18$ & $60.66$  & $18.73(26)$ & $17.48(10)$ & $16.89(06)$ & $16.20(06)$ & KAIT \\
$56523.21$ & $63.68$  & $18.70(06)$ & $17.49(03)$ & $16.88(02)$ & $16.27(01)$ & Nickel \\
$56528.18$ & $68.62$  & $19.07(40)$ & $17.75(15)$ & $17.00(08)$ & $16.40(07)$ & KAIT \\
$56530.17$ & $70.61$  & $18.60(25)$ & $17.64(10)$ & $17.12(06)$ & $16.36(05)$ & KAIT \\
$56531.19$ & $71.62$  & $18.97(05)$ & $17.65(02)$ & $17.07(01)$ & $16.43(01)$ & Nickel \\
$56532.16$ & $72.60$  & $19.08(29)$ & $17.75(12)$ & $17.12(06)$ & $16.42(07)$ & KAIT \\
$56534.16$ & $74.58$  & $18.73(31)$ & $17.69(10)$ & $17.07(07)$ & $16.40(05)$ & KAIT \\
$56540.16$ & $80.55$  & $18.92(27)$ & $17.79(12)$ & $17.30(07)$ & $16.47(05)$ & KAIT \\
$56544.16$ & $84.53$  & $19.24(48)$ & $17.78(13)$ & $17.33(08)$ & $16.62(08)$ & KAIT \\
$56675.58$ & $215.36$ & $20.60(17)$ & $20.94(17)$ & $19.75(06)$ & $19.16(06)$ & Nickel \\
$56689.58$ & $229.30$ & $>21.45$    &  $\cdots$   & $\cdots$    & $\cdots$    & Nickel \\
$56713.56$ & $253.18$ & $>21.61$    &  $\cdots$   & $\cdots$    & $\cdots$    & Nickel \\
$56724.53$ & $264.11$ & $>21.54$    &  $\cdots$   & $20.51(09)$ & $\cdots$    & Nickel \\
$56738.49$ & $278.00$ & $>21.41$    &  $\cdots$   & $20.69(14)$ & $20.48(19)$ & Nickel   
\enddata 
\tablecomments{Uncertainties in parentheses given in units of 0.01\,mag.} 
\tablenotetext{a}{Rest-frame phase after explosion (JD\,$=2456459.24$).}
\end{deluxetable*} 

\subsection{Distance and Extinction}
\label{sec:dist}

\noindent In the current analysis we adopted a distance to
NGC~5806 of $25.8\pm2.3$\,Mpc as provided by the NASA/IPAC
Extragalactic Database (NED). 
This value 
is similar to that used by \citet{Bersten14}, and about 10\%
(1$\sigma$) greater than the 22.5\,Mpc extensively adopted in
the literature on iPTF13bvn.  

Milky-Way reddening in the direction to iPTF13bvn is
$E(B-V)_{\mathrm{MW}}=0.045$\,mag \citep{Schlafly11}. 
\citet{Cao13} found a total color excess of
$E(B-V)_{\mathrm{tot}}=0.07$\,mag by measuring the equivalent width of
Na~I~D lines in the SN spectrum and using the relation by
\citet{Poznanski12}. However, \citet{Phillips13} pointed out that such
a relation underestimates the uncertainty in the extinction. Using 
observed colors in comparison with a sample of stripped-envelope SNe,
\citet{Bersten14} found a larger value of 
$E(B-V)_{\mathrm{tot}}=0.21\pm0.03$\,mag. \citet{Srivastav14} favored
this larger reddening value by comparing the $V-R$ color with the
calibration of intrinsic color provided by \citet{Drout11}. We thus
adopted the larger reddening value and
computed extinction in all bands using the reddening law of
\citet{Cardelli89} with a standard coefficient of $R_V=3.1$. Where
indicated, we also considered the shorter distance and lower 
extinction values from \citet{Cao13}.

\section{THE PROGENITOR OF iPTF13bvn}
\label{sec:prog}

\noindent With the magnitudes and detection limit listed in
Table~\ref{tab:mag}, we constructed the spectral energy
distribution (SED) of the pre- and post-SN
source, as shown in the right panel of Figure~\ref{fig:lcssed}. 
The source at the SN location faded below the pre-explosion level
in 2015. The decrease is significant in the 
F435W/F438W and F555W bands (by $\sim3\sigma$ and $\sim4\sigma$,
respectively), and marginal in F814W. To test this observation, we performed
pre-SN minus post-SN image subtractions, using AstroDrizzle
to register and degrade the post-SN images to match the pre-SN
images. Then we scaled the flux and subtracted with standard IRAF
routines. As shown in 
Figure~\ref{fig:img}, the subtractions leave detectable sources at the
SN location in F435W/F438W and F555W, and only noise in
F814W. We performed aperture photometry of the residual object using
DAOPHOT, which led to the magnitudes and upper limit listed in
Table~\ref{tab:mag}. It is reassuring that the flux in the subtracted
images closely matches the subtraction of fluxes measured in the 2005
and 2015 images.

Assuming that the flux decrease is not caused by large amounts of dust
formed in the SN ejecta (see Section~\ref{sec:casa}),
this result confirms the first association of a SN~Ib with its
progenitor object.  
However, deriving a conclusion about the progenitor nature  
requires some interpretation of the new measurements. 
In principle, the flux in 2015 could be produced by any combination of
the following: the fading SN, a light echo, or
an underlying object or population related (or not) to the SN.
If what we detected is purely the SN, then the pre-SN
source is the progenitor itself. In the opposite extreme, if the SN 
makes a negligible contribution, then the progenitor is revealed in
the subtraction of pre-SN minus post-SN images (assuming no variability
of the environment). Any intermediate situation is theoretically
possible. In the following we shall analyze both extreme cases. 

\subsection{Case SN: The Fading SN}
\label{sec:casa}

\noindent In order to tell if the detected flux in 2015 was caused by the 
SN, we compared it with available data on similar events. Unfortunately, there
are no observations of other SNe~Ib at such a late phase. The only
published multiband light curves of stripped-envelope SNe that extend
over 700\,days are those of the Type~IIb SN~1993J and SN~2011dh. However,
these SNe may be affected by stronger emission from shock interaction
than iPTF13bvn. Indeed, after 500\,days  
SN~1993J showed evidence of prominent interaction emission that flattened
the optical light curves \citep{Zhang04}. 
In contrast, SN~2011dh appears to have been relatively free of
interaction \citep[e.g.,
  see][]{Maeda14,Jerkstrand15}. To test this we performed a simple, 
one-zone radioactive deposition calculation \citep{Maeda03},
assuming full positron trapping and $\gamma$-ray optical depth based
on the ejected mass and explosion energy given by
\citet{Bersten12}. The resulting optical emission is enough to
account for the late-time observations of SN~2011dh without the need
to invoke strong interaction. 

Figure~\ref{fig:lcssed} (left panel) shows the $BVRr'I$ light curves
of SN~2011dh  
\citep{Ergon14,Ergon15,VanDyk13}, assuming $d=7.8$\,Mpc and $A_V=0.1$\,mag
\citep{Folatelli14}, and scaled to the
distance and extinction of iPTF13bvn. We extrapolated these light
curves to 740\,days (dotted lines), assuming an
average decline rate of $0.007\pm0.002$\,mag/day measured between 600
and 700\,days. The extrapolation uncertainty was summed in quadrature
with the uncertainty of the latest observed points.
We decided not to include later observations of
SN~2011dh published by \citet{Maund15} because they were obtained at
$\sim1160$ days (i.e., $>400$\,days after the epoch of our
observations), when different emission mechanisms may
dominate. The right panel of Figure~\ref{fig:lcssed} shows that the
resulting SED of SN~2011dh is very similar to that of iPTF13bvn, only
slightly fainter in the F435W/F438W band, suggesting
that our new images of iPTF13bvn reveal the fading SN.

We also computed spectrophotometry with the
spectrum of iPTF13bvn obtained at 306\,days by
\citet{Kuncarayakti15}. The synthetic fluxes are shown in the right
panel of Figure~\ref{fig:lcssed}, scaled down to
reproduce the observed magnitude of $\mathrm{F555W}=26.72$ at 740\,days. 
The overall SED shape is similar to our measurements at 740\,days, 
which may imply that the spectrum did not evolve
significantly. Along with the similarity with SN~2011dh, this suggests
the absence of strong interaction or large dust formation, unless both
effects canceled each other. The small ``excess'' in the F438W band
seen at 740\,days may hint the presence of a hot companion
star, as shown in the figure.

Assuming we detected the fading SN, the pre-SN flux could be
attributed primarily to the progenitor. The pre-explosion SED has been
compared with possible progenitor models in several previous articles,
as described in Section~\ref{sec:intro}. Here we revisit the
progenitor nature using our own pre-explosion photometry
and our revised distance and reddening (Section~\ref{sec:dist}).  

The absolute magnitudes of the progenitor object would be
$M_{\mathrm{F435W}}=-6.95\pm0.28$\,mag, 
$M_{\mathrm{F555W}}=-6.69\pm0.26$\,mag, and 
$M_{\mathrm{F814W}}=-6.61\pm0.24$\,mag. The uncertainties were
derived by summing in quadrature those in apparent
magnitude, extinction, and distance. Intrinsic colors
would be $(M_{\mathrm{F435W}}-M_{\mathrm{F555W}})=-0.26\pm0.19$\,mag and 
$(M_{\mathrm{F555W}}-M_{\mathrm{F814W}})=-0.08\pm0.18$\,mag.

We compared the corrected photometry with available models of single
and binary massive stars, as shown in the color-magnitude and color-color
diagrams of Figure~\ref{fig:colmag}. 
The rotating models of 20\,M$_\odot$ and 28\,M$_\odot$ of
\citet{Groh13a} (green and blue triangles) are in good agreement with
the data. The 
pre-SN objects predicted by these 
two models are a luminous blue variable (LBV) and a WR star of type
WN10-11, respectively. Their respective final masses are 7.1\,M$_\odot$
and 10.8\,M$_\odot$. The mass of the latter object is inconsistent
with the analysis of the 
early-time SN light curve. While the LBV model is marginally
compatible with the light-curve analysis, its final structure
contains significant amounts of hydrogen and would thus produce a Type
IIL or IIb~SN. We note, however, that there have been claims of the
presence of H$\alpha$, albeit weak, in the spectrum of iPTF13bvn 
\citep[][although see \citealt{Cao13} and \citealt{Srivastav14}]{Kuncarayakti15,Reilly16}.

We also compared with massive binary models published by
\citet{Kim15} and E15. Although
  E15 models with initial primary masses below
  13\,M$_\odot$ produce faint companion stars in agreement with the
constraint in F225W, those systems are slightly less luminous or
redder than our pre-SN measurements. Only those models by
\citet{Kim15} that included the most massive companion 
star (35\,M$_\odot$) were able to account for the large optical
luminosity. However, these companions would be too bright in F225W to
comply with our detection limit. As shown in the right panel of
Figure~\ref{fig:lcssed}, only their least massive (20\,M$_\odot$) 
companion is allowed by our F225W constraint. We note that such
  would be the limit of the companion mass at the moment of the SN
  explosion. Binary evolution models have shown that accreting
  companions evolve upward in the Hertzprung-Russell diagram,
  remaining near the zero-age main sequence (ZAMS). However, deriving the
  initial binary configuration would require detailed evolutionary
  calculations that would only be justified once we confirm the nature
  of the source observed in 2015 with further observations.

For comparison, Figure~\ref{fig:colmag} also shows the results
with the shorter distance and lower extinction (see
Section~\ref{sec:dist}). The resulting values, indicated with ``low
$A_V$'' in Figure~\ref{fig:colmag}, are  
$M_{\mathrm{F435W}}=-6.06\pm0.28$\,mag, 
$M_{\mathrm{F555W}}=-5.92\pm0.26$\,mag, and 
$M_{\mathrm{F814W}}=-6.06\pm0.24$\,mag. 
The agreement with the models by \citet{Groh13a} and \citet{Kim15}
is worse than with our preferred distance and extinction. Among the binary
systems by E15, those with initial masses of
9\&8.1\,M$_\odot$ (and $\log(a/{\mathrm{R}}_\odot)=2.25$), and
10\&7\,M$_\odot$ agree within $1\sigma$ of the low-$A_V$ photometry.

\subsection{Case Env: The Environment}
\label{sec:casb}

\noindent We now consider the
case that the SN emission became negligible at 740\,days. This requires
removing the condition of full trapping of radioactive
positrons and neglecting any contributions from possible weak
interaction or light echo, or assuming much obscuration by dust. By
linearly extrapolating the light curves from $\sim200$ days, 
we find the SN flux would contribute 
$<1$\% of the observed flux in F438W and F555W, and
$<6$\% in F814W. This would mean that the new observations revealed the
SN environment. The 740\,days SED is roughly compatible with an early-A
type supergiant star or a young star cluster. At
the distance of NGC~5806, the PSF radius comprises several
parsecs; thus, the source may or may not be physically 
related to the SN progenitor. Our data rule out a
bloated red supergiant companion star as proposed by \citet{Hirai15};
only their hottest companion (model c0.5) would be allowed.

\begin{figure*}[htpb] 
\plottwo{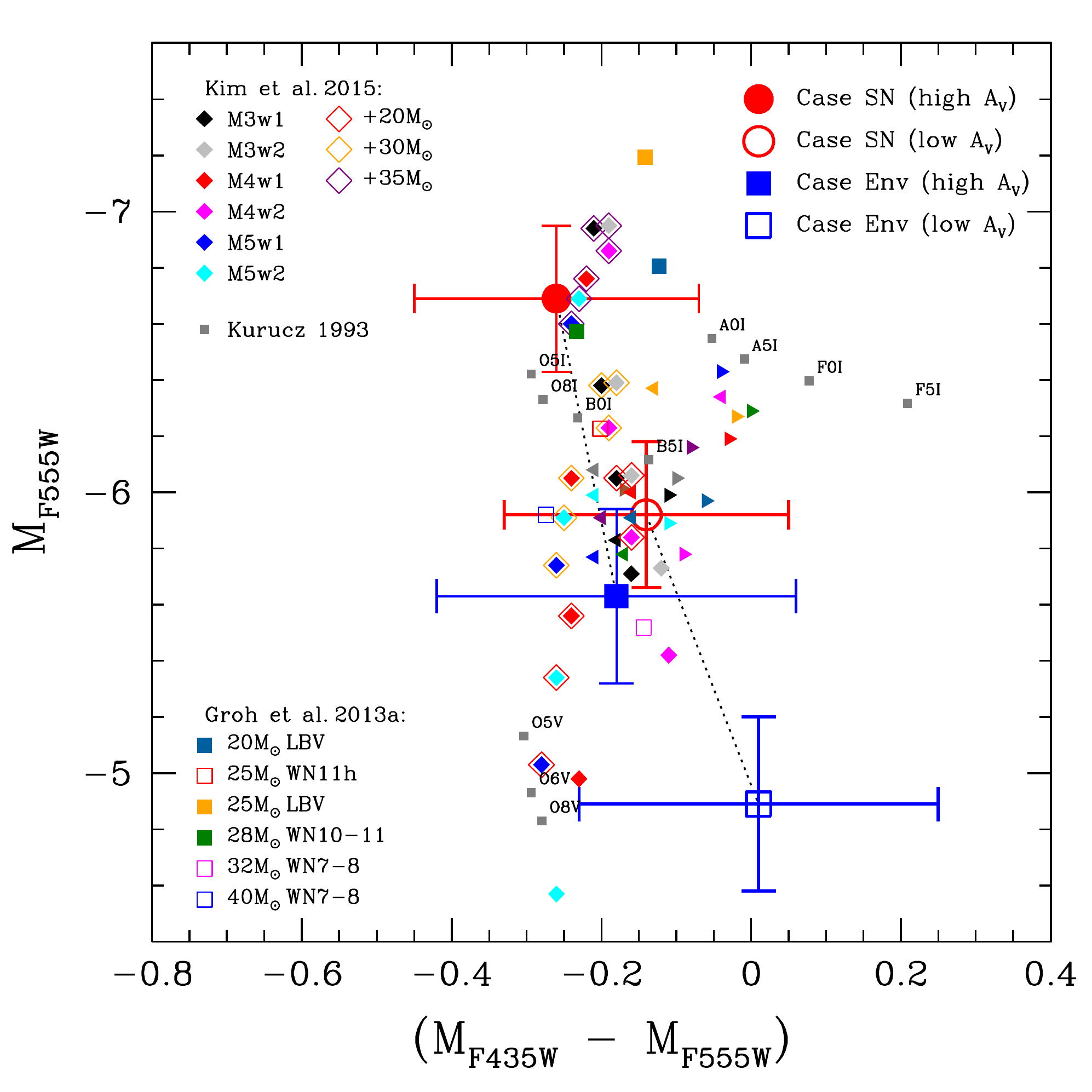}{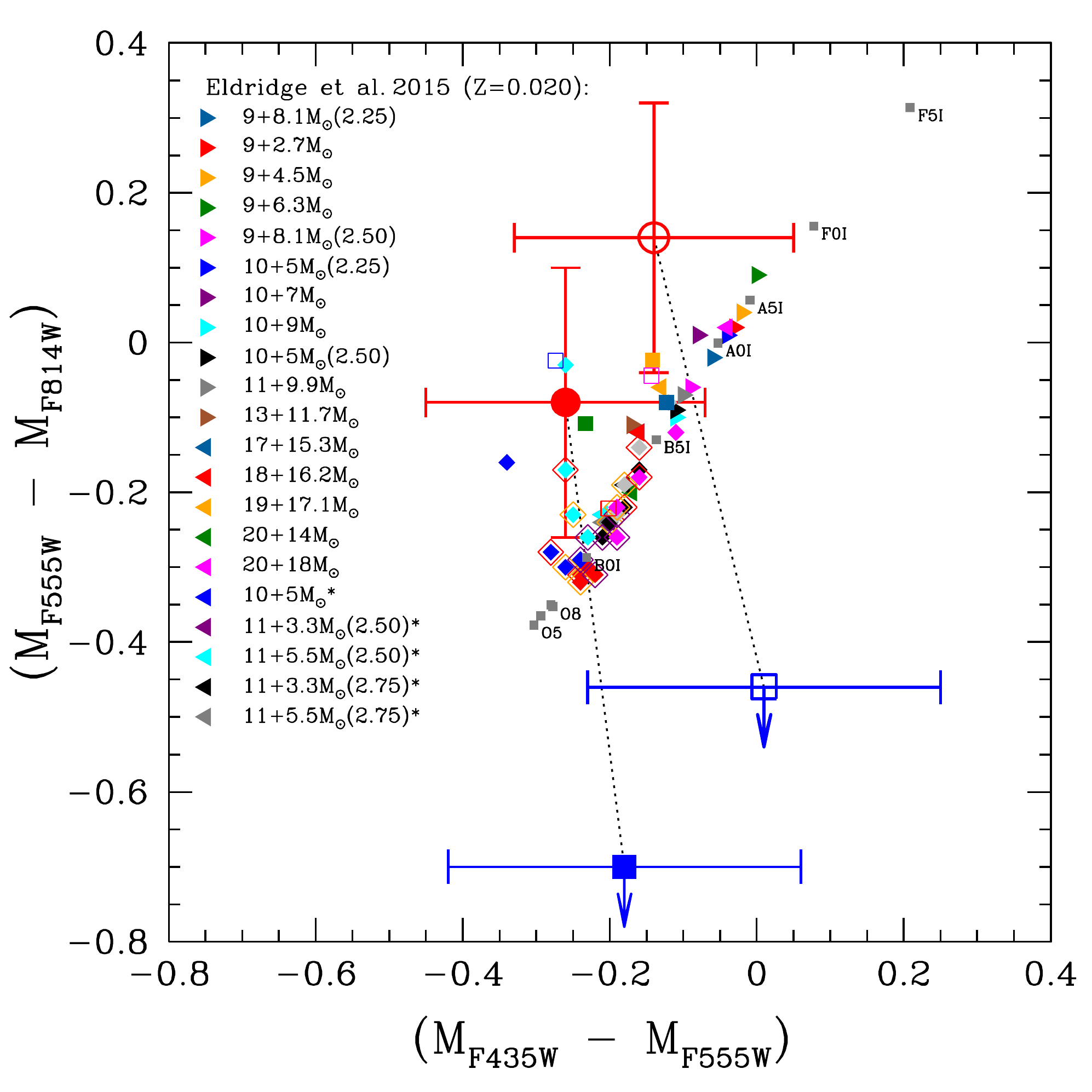}
\caption{Color-magnitude and color-color diagrams showing
  the location of the two extreme progenitor alternatives (Case SN and
  Case Env) discussed in Section~\ref{sec:prog}, compared with previously
  proposed progenitor models \citep{Groh13a,Kim15,Eldridge15} and stellar
  atmosphere models \citep{Kurucz93}. Only solar-metallicity models by
  E15 are shown, with labels indicating the initial
  masses of both components. Values in parentheses give the initial
  orbital separation ($\log(a/{\mathrm{R}}_\odot$), as in Table~1 of
  E15) for systems of equal masses. Dotted lines join
  both extreme cases. ``High $A_V$'' indicates our
  assumed distance and extinction (Section~\ref{sec:dist}), and ``low
  $A_V$'' is for values adopted by \citet{Cao13}. Arrows in the right
  panel indicate upper limits for Case Env.
  \label{fig:colmag}}
\end{figure*}

Assuming no variability, the pre-SN minus post-SN
images would reveal the progenitor object, the exploding star
itself, devoid of any binary companion or associated population. 
The absolute magnitudes would be
$M_{\mathrm{F435W}}=-5.81\pm0.27$\,mag, 
$M_{\mathrm{F555W}}=-5.63\pm0.31$\,mag, and  
$M_{\mathrm{F814W}}>-4.93$\,mag. If we consider that up to 6\% of the
flux in F814W could be due to the SN, the progenitor would have
$M_{\mathrm{F814W}}\gtrsim-5.0$\,mag. If we consider only the 
F435W and F555W bands, we can find compatible progenitor models
(Figure~\ref{fig:colmag}, left panel). However, when including the
limit in F814W (right panel), the resulting color of  
$(M_{\mathrm{F555W}}-M_{\mathrm{F814W}})\lesssim-0.6$\,mag
excludes all of the comparison models. 
Even if we adopted the shorter distance and lower extinction
(Section~\ref{sec:dist}), the color 
$(M_{\mathrm{F555W}}-M_{\mathrm{F814W}})\lesssim-0.4$\,mag would still
be beyond the range of known models, although close to the locus of
O-type stars. At the same time, the
$M_{\mathrm{F435W}}-M_{\mathrm{F555W}}$ color worsens. We
conclude that the true nature of the progenitor is likely between
the two cases, and closer to Case SN (Section~\ref{sec:casa}) than to
this other extreme. 

\section{CONCLUSIONS}
\label{sec:concl}

\noindent Our new {\em HST} images reveal a decrease in flux relative to
the pre-explosion source proposed as the progenitor of iPTF13bvn by
\citet{Cao13}. Assuming this decrease is not caused by newly formed dust
in the ejecta, it confirms the first progenitor identification of a
SN~Ib. 

With the currently available information it is not possible to provide a
definitive characterization of the progenitor. 
We present some evidence that the flux in the new images was
mostly due to the fading SN. In fact, if we assume a negligible
contribution from the SN, the derived progenitor
$M_{\mathrm{F555W}}-M_{\mathrm{F814W}}$ color would be incompatible
with known stellar models.

If what we detected was indeed the SN, then
most of the pre-SN flux would be due to the progenitor. Previously proposed
progenitor models, either single or binary, need to be revised to
account for these new data. For binary progenitors, our detection
limit in F225W constrains any hot companion to be less luminous
than a late-O main-sequence star, with $\lesssim20$\,M$_\odot$ at
  the moment of the SN explosion, assuming it is not heavily obscured
by dust. Further observations are required to assess the exact
contribution from the SN, and thus to disentangle the progenitor
nature.  

\acknowledgments
 
This research is supported by grants GO-13684, GO-13822, and AR-14295
from the Space Telescope Science Institute (STScI), which is operated
by the Association of Universities for Research in Astronomy (AURA),
Inc., under NASA contract NAS5-26555.
A.V.F.'s group is also grateful for funding through NSF grant AST-1211916,
the TABASGO Foundation (KAIT and research support), the Sylvia \&
Jim Katzman Foundation, Clark and Sharon Winslow, and 
the Christopher R. Redlich Fund.
This research is supported by the 
World Premier International Research Center Initiative 
(WPI) Initiative MEXT (Japan), 
the Japan Society for the Promotion of Science (JSPS) KAKENHI
Grants 26800100 (K.M.) 23224004, and 26400222 (K.N.), and by the JSPS
Open Partnership Bilateral Joint 
Research Project between Japan and Chile (K.M.).
M.H., G.P., and H.K. acknowledge support from the
Millennium Institute of Astrophysics (MAS; Programa Iniciativa
Cient\'ifica Milenio del Ministerio de Econom\'ia, Fomento y Turismo
de Chile, grant IC120009). H.K. also acknowledges FONDECYT grant
3140563. N.E.R. is supported by PRIN-INAF 2014 with the project
``Transient Universe: unveiling new types of stellar explosions with
PESSTO''.  
R.J.F. acknowledges support from NSF grant AST--1518052
and the Alfred P. Sloan Foundation. A.A.M. acknowledges support by
NASA from a Hubble Fellowship grant: HST-HF-51325.01, awarded by
STScI, operated by AURA, Inc., for NASA, under contract NAS
5-26555. Part of the research was carried out at the Jet Propulsion
Laboratory, California Institute of Technology, under a contract with
NASA. We also thank U.C. Berkeley students Samantha Stegman, Sahana Kumar, 
Kevin Hayakawa, Kyle McAllister, Carolina Gould, Goni Halevy, Xianggao
Wang, Anthony Khodanian, Minkyu Kim, Heechan Yuk, 
Andrew Bigley, Michael Hyland and Timothy Ross for their effort in
taking Lick/Nickel data. Research at Lick Observatory is partially
supported by a generous gift from Google.

\end{document}